# Analyzing Image-based Political Propaganda in Referendum Campaigns: From Elements to Strategies

Ming-Hung Wang[1*], Wei-Yang Chang[2], Kuan-Hung Kuo[1] and Kuo-Yu Tsai[2]

[1]Department of Computer Science and Information Engineering, National Chung Cheng University, Chiayi, Taiwan

[2]Department of Information Engineering and Computer Science, Feng Chia University, Taichung, Taiwan

## Abstract

With the increasing popularity of social network services, paradigm-shifting has occurred in political communication. Politicians, candidates, and political organizations establish their fan pages to interact with online citizens. Initially, they publish text-only content on sites; then, they create multimedia content such as photos, images, and videos to approach more people. This paper takes a first look at image-based political propaganda during a national referendum in Taiwan. Unlike elections, a referendum is a vote on policies. We investigated more than 2,000 images posted on Facebook by the two major parties to understand the elements of images and the strategies of political organizations. In addition, we studied the data collection's textual content, objects, and colors. The results suggest the aspects of propaganda materials vary with different political organizations. However, the coloring strategies are similar, using representative colors for consolidation and the opponent's colors for attacks.

## Introduction

Social network services have triggered a paradigm shift in political communication. Not only politicians but also political organizations established their fan pages to interact with online citizens on social network platforms such as Facebook and Twitter. Moreover, they use social services to disseminate their views and advertise themselves during elections. These channels enable candidates to approach voters who may not be exposed to TV and newspaper. Many studies have analyzed the textual content published on Facebook and Twitter to understand how political entities use social platforms to create propaganda [1], [2]. Meanwhile, the formats of propaganda and the platforms where propaganda was conducted also evolve with time. For example, political advertisement not only appears on Twitter but also on multimedia intensive platforms such as Instagram and Tiktok [3], [4]. Also, the text messages have been embedded with photos and video clips to attract more users. Thus, though massive research has investigated how social media act in election campaigns in recent years, few studies about how political entities create multimedia content and use it in online political propaganda, especially in a national referendum.

Therefore, we provide a very first look at online political propaganda during a national referendum in this work. Three major research questions investigated are described as follows:

1. What are the elements (texts and objects) in the image-based propaganda materials?

2. What are the color usage of the government and the opposition during a referendum?

3. What are the differences in images used for various propaganda purposes (e.g., attacking/consolidation)?

To answer these questions, we study the election campaign on Facebook for a national referendum in Taiwan in 2021. We select a referendum as the target because, unlike elections are votes on candidates, the referendum is on a proposal or a policy; thus, we are interested in how the format and content of online propaganda will vary in the referendum compared to elections.

We collected more than 10 thousand images published by the two major parties, Democratic Progressive Party (DPP, the ruling party) and Kuomintang (KMT), on Facebook during an 8-month-long observation before the referendum. We filtered out the top 1000 images of DPP and KMT based on the number of reaction emojis expressed by the users. We performed text recognition, object extraction, and color quantification on the dataset; a detailed comparison between the two parties is made and discussed in this work. Overall, the main contributions of this work are three-fold:

1. Political entities publish propaganda text on Facebook along with multiple images.

2. The infographic has been an essential format for political communication during a referendum.

3. When consolidating supporters, political parties like to use their representative color for creating images.

4. When attacking the opposition, political parties tend to use the opposition's representative color more to design the material.

This paper is organized as follows: In Section Related Works, we introduce related works about political propaganda and the background of the referendum studied in this paper. In Section Methods, we present the detail of data collection and extraction methods for texts, objects, and colors. In Section Results, we show a thorough analysis of the texts, objects, and color usage in the images published by the two major parties. In Section Analysis, we demonstrate the in-depth analysis of propaganda strategies. Finally, the conclusion and discussion of this work are depicted in Section Conclusion.

## Related Works

Harold D. Lasswell [5] first proposed a theory of political propaganda in 1927 and defined political propaganda as "the management of collective attitudes by the manipulation of significant symbols." Since then, political propaganda has become a primary way to deliver messages, proposals, and thoughts from political elites to the general public. In 1997, R.A. Nelson [6] described political propaganda usually spread through the dissemination of unilaterally controlled messages. These messages attempt to influence a particular target's emotions, attitudes, beliefs, and behaviors were spread through large-scale, direct media channels. In the modern era, political propaganda has been redefined due to the development of the Internet. Marlin, Randal [7] defined propaganda as "the organized attempt, through communication, to affect belief or action or inculcate attitudes in a large audience in ways that circumvent an individual's adequately informed, rational, reflective judgment." Étienne Brown [8] considers the various means used by propagandists and categorizes three ideal types of propaganda: affective propaganda, conative propaganda, and cognitive propaganda. In 2018, Jowett et al. [9] defined political propaganda as a deliberate, systematic attempt to form ideas, manipulate perceptions, and guide behavior to achieve the desired response. With the rapid development of social network services, channels for people to receive information have extended from offline media such as newspapers and TV to online platforms like Facebook and Twitter. We introduce the works related to political propaganda on social platforms and discuss the strength and necessity of our study.

## Political Propaganda on Social Platforms

Social network services have enabled users to receive information and share their opinions. Meanwhile, politicians and organizations have started establishing their fan pages to speak with the general public and interact with their audience. However, past studies have confirmed that citizens are far more interested in the characteristics of the party or candidate than in specific issues [10]–[12]. Due to these findings, some of the candidates or parties will like to post more photos and positive comments to manage their image, which is an idea of propaganda. To study the impact of online propaganda, Woolley et al. [13] investigated how political parties spread their ideas through bot accounts on social media and their effects during major political events. Ferrara et al. [14] took the U.S. as an example to study the impact of online political propaganda on Twitter; they classified bot accounts as right and left-leaning to understand how political propaganda affects the presidential election.

As political elites are getting engaged in social platforms to communicate with citizens, scholars have investigated content published and replied to by political entities to examine how social communication can influence the public. Brady et al. [15] studied the gauge influence of political elites on social media. They found a "moral contagion" effect, which refers to "elites' use of moral-emotional language was robustly associated with increased message diffusion." To analyze the topic and emotion of political propaganda, Daniel Taninecz Miller [16] studied Russian propaganda on Twitter before, during, and after the 2016 U.S. presidential election. The results show that Russian sponsored tweets have distinct attitudes towards different presidential candidates.

Political propaganda is a common approach in political communication to advocate specific proposals or influence people's political tendencies, texts, photos, or even videos. Before understanding the political propaganda in images, most research focuses on detecting political propaganda through text mining or bot account detection [17]–[20]. In the study of image propaganda, we can split the survey into different classes, such as candidates or party's images, events, or even ideological. Mattes, Kyle et al. [21] studied how the candidate images will affect the election results through crowdsourcing, collecting the impression netizens felt from the photos. On the side of detecting events through social media images, Won, Donghyeon et al. [22] develop a visual model to recognize protesters and estimate the level of perceived violence in a snap. Aim to detect the protest event and statistics on perceived violence in different areas. Zhang, Han [23] tried to identify collective action events through text and image data from Sina Weibo, which is China's biggest microblogging platform. They classify text by Long short-term memory (LSTM) and extract image features from Convolutional Neural Network (CNN) through combined these two results to identify the events in real life. In research on political parties or ideological fields, Xi, Nan et al. [24] tried to understand how legislators spread their political ideology through social media images. They detect the scene and the facial expression to analyze the difference between Democratic or Republican politicians. Amogh Joshi et al. [25] examining similar and ideologically correlated imagery of political communication through clustering and deep learning. They collected images of US congress politicians from Twitter's REST API and classified images correlated with liberal or conservative ideology.

According to the above literature review, we figured out that most research on images focuses on object detection, human facial expression, or extracting image features [22]–[25]. Inspired by these past studies, we apply the research in Taiwan politics and extend the research structure. Unlike text-based political propaganda research, we extract the text content in images to observe any systematic text design used for image propaganda. Previous research detects objects and analyzes them through known correspondence; in contrast, we try to detect objects and observe the correspondence between different political parties. Most importantly, we break through the analysis on images, combining quantifying and analyzing the color in pictures with object detection, hoping to figure out any pattern they design propaganda images.

## Taiwanese Politics and the National Referendum in 2021

Taiwan politics have long been polarized since the 1980s. The two major parties, DPP (the current ruling party) and KMT, have won elections and held a majority of elected seats in the parliament alternately in the last 20 years [26]–[29]. The two parties have different attitudes toward the relationships between China and the U.S.; their thoughts about the national identity of Taiwan are also polarized online, and offline [30]–[33] . In 2021, a national referendum was held in Taiwan to vote on four questions, including 1) activating the Fourth Nuclear Power Plant in Gongliao District, New Taipei City, 2) banning imports of pork containing ractopamine, 3) holding referendums alongside nationwide elections, and 4) relocating a Natural Gas Terminal to protect algal reefs off Taoyuan Guanyin District. From the results of the referendum, the four proposals are all rejected. Among these proposals, banning U.S. pork imports is the most controversial one and has raised domestic and international attention. In 2020, president Tsai Ing-wen announced that the government would lift the ban on U.S. pork products in 2021. After that, politician's from KMT proposed a referendum to impose restrictions on the import of pork containing ractopamine due to concerns about food safety. However, the government claims the regulation could potentially damage the relationships between Taiwan and the U.S.; also, it would affect the progress of Taiwan in joining international trade agreements. The outcome is that 51.2% voted against the ban while 48.8% voted for it.

## Data & Methodology

The overall process of our analysis scheme was described in Figure 1. There were five components in the process, including 1) data collection, 2) data preprocessing, 3) image categorization, 4) object detection, and 5) color quantification. Finally, we visualized the results and performed an in-depth analysis of them.

### Data Collection

We collected our dataset on Facebook from April 2021 to December 2021, an 8-month-long observation before the referendum. We crawled a total of 103,845 posts, of which post content was related to the four referendum questions, as described in Section 2.2. Among these posts, only 40,953 (39%) included at least a photo in a post. To investigate the propaganda conducted by the two major parties, we labeled each post publisher according to their political affiliation. Only images published by the politicians or affiliations of the two major parties are included in our dataset. A summary of our dataset is shown in Table 1.

*Table 1 The information about dataset.*

| Item | Info |
|---|---|
| Data collect | 2021/04/01~2021/12/18 |
| fan page | 423 |
| # post | 103,845 |
| # post about referendum | 6,854 |
| # image about referendum | 9,351 |

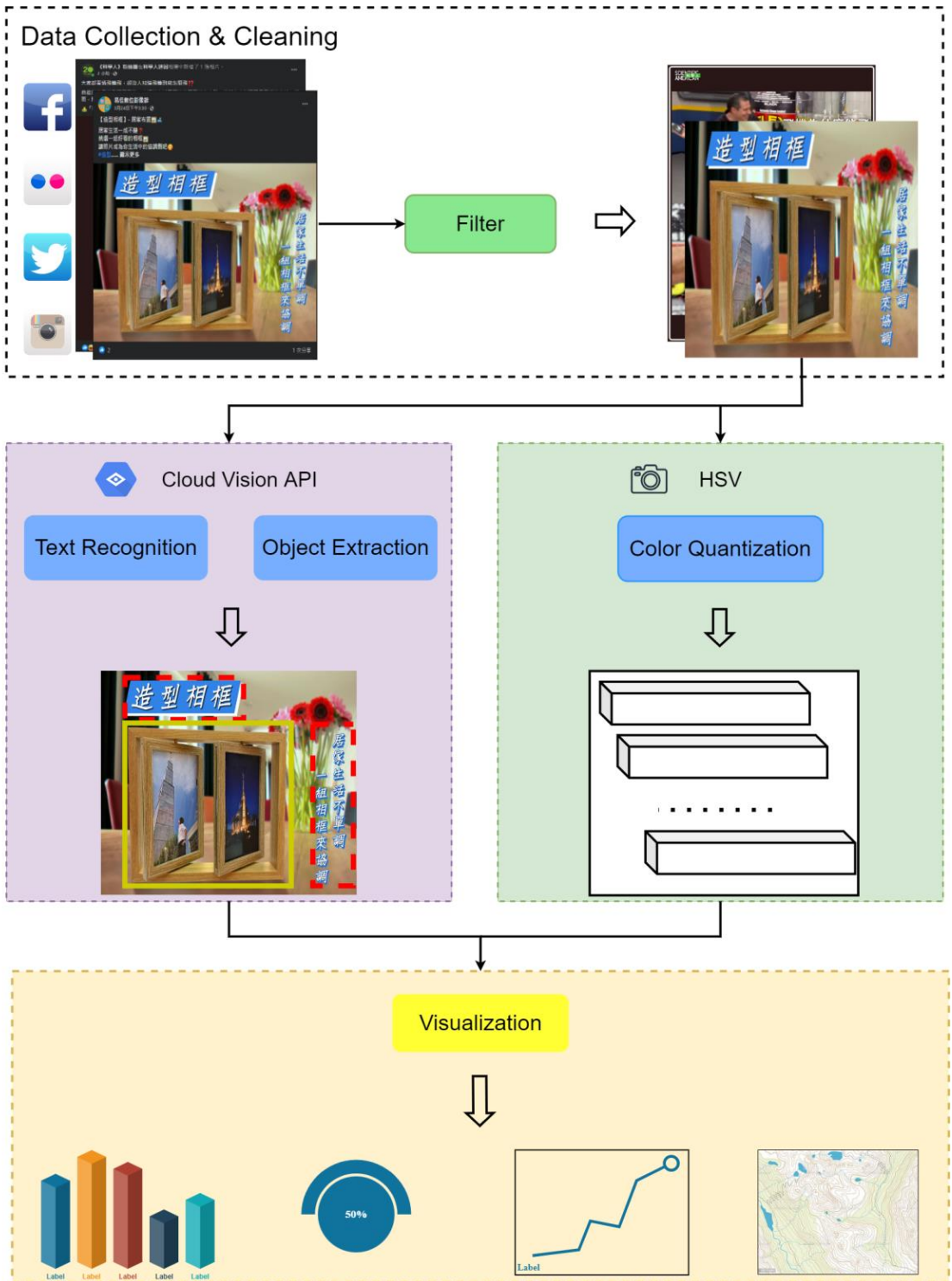

*Figure 1 The analysis architecture in this paper.*

## Text Recognition

To investigate the textual content of images, we extracted texts through optical character recognition provided by Google Cloud Vision API [1]. We performed the word segmentation on the text extracted from images using the Chinese Knowledge and Information Processing (CKIP) library [2].

## Image Categorization

To figure out the differences in presentations and strategies during political propaganda, we categorized our data collection into two classes, infographic and slogan images. The categorization guide is described as follows:

- **Infographic**
  Infographic referes to an image used as signs, reports, and technical documents. The formats include but not limited to timelines, maps, data visualizations, flow charts, and statistic charts.

- **Non-infographic**
  An image includes slogans without any statistic charts or data visualizations.

Some of the image type we described above was shown in Figure 2. In addition to the above two image types, a large portion of our collection was shot by photographers, such as photos of large-scale outdoor campaigns with crowds.

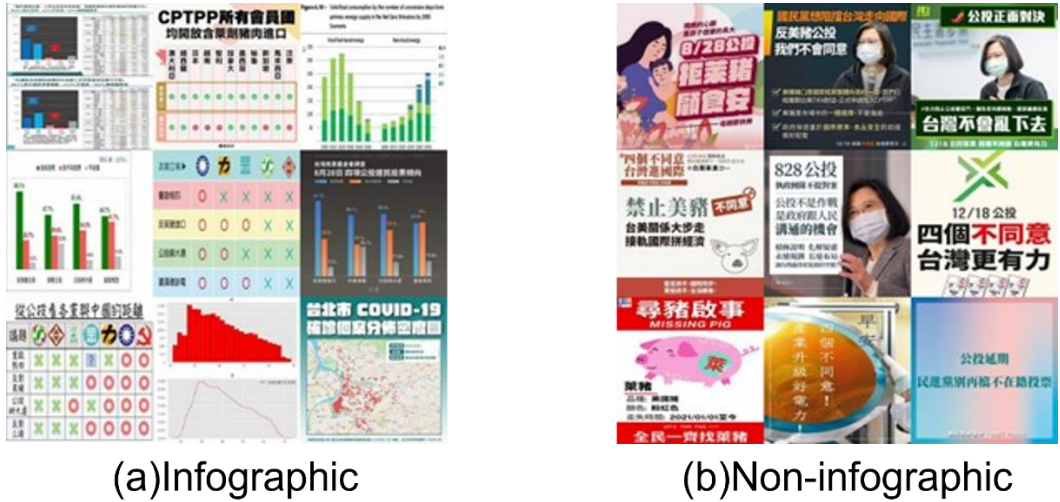

*Figure 2 Example for infographic and non-infographic images.*

---

[1] https://cloud.google.com/vision

[2] https://ckip.iis.sinica.edu.tw/

## Object Extraction

To extend the previous studies on textual content analysis about political propaganda [34]–[36], we investigated the elements in our data collection. We aimed to realize how different political organizations create their propaganda materials to symbolize their idea and proposal. We extracted the objects inside images through Google Vision API. We processed the derived labels from the platform and categorized the tags into three subcategories, including clothing, objects, and background. We analyzed the difference between political entities with polarized stances.

## Color Quantification

According to many previous researches, color could be an important factor in psychological effects and influence decision-making according to many previous researches [37], [38]. To realize the impact of colors in image-based political communication, we quantified the color distribution of images and investigated 1) how organizations use color to produce their propaganda material? And 2) are there different strategies in using colors for other purposes (attacking the opposite or consolidating supporters)?

We performed a color quantification process on our dataset to understand the color composition of images. To find the dominant colors of an image, we clustered pixels in terms of RGB and HSV through k-means, where $k = 5$ in our setting. Then, we chose the center of each cluster and calculated the percentage this cluster occupies to extract the five dominant colors of every image.

- **RGB (RGB color model):**
  RGB is an additive color model for representing a color via red, green, and blue. In the system, red, green, and blue index are all set in range $0 \sim 255$, where (0,0,0) represent black and (255,255,255) represent white.

- **HSV (Hue, Saturation, and Value):**
  HSV is used to represent colors through a cylindrical coordinate system. Hue is a basic property for colors, such as red, yellow, and green. In a cylindrical coordinate system, the hue is represented as an angle in $0° \sim 360°$, but in the python OpenCV module, it is represented as $0 \sim 180$. Saturation is the purity value of colors, which means colors with a low saturation value will turn gray. We set saturation in the range of $0 \sim 255$ in our program. Value can also be named brightness. A color with a low brightness value would turn black.

Figure 3 demonstrates the dominant colors extracted by $k = 5$ and $k = 10$. Though setting $k = 10$ can help us derive a more detailed description of color distribution, $k = 5$ can reserve the most dominant colors with less computational effort. We visualized and analyzed the color usage for different parties, different categories of images, and different image purposes from the results.

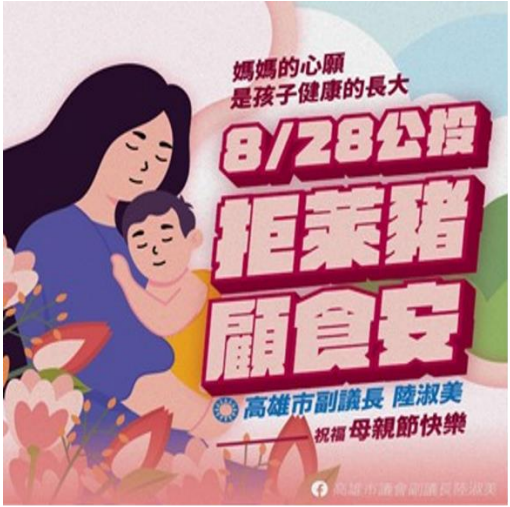
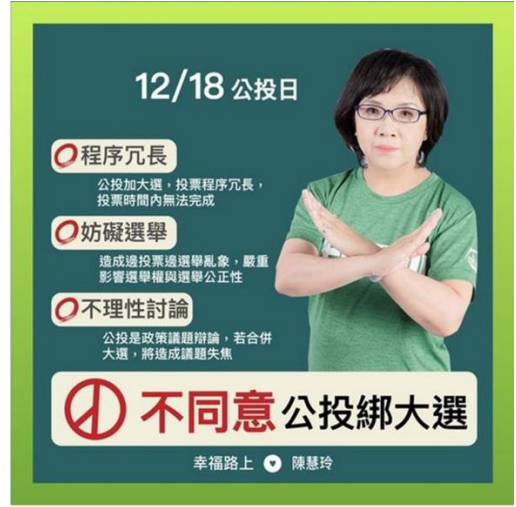

(a)KMT(pan-blue)                          (b)DPP(pan-green)

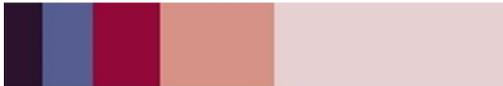
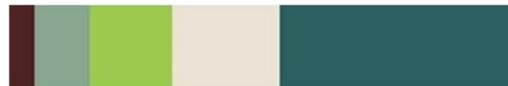

(c)KMT(pan-blue) $k = 5$              (d)DPP(pan-green) $k = 5$

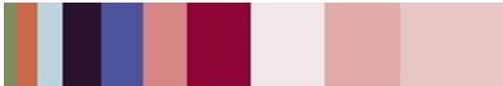
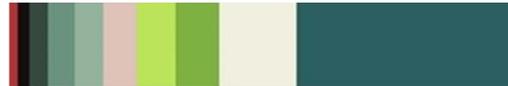

(e)KMT(pan-blue) $k = 10$            (f)DPP(pan-green) $k = 10$

*Figure 3 The result of k-means clustering at k=5 and k=10.*

## Experiment and Result

From the data collection, we selected the top 1,000 images from pan-blue and pan-green fan pages on Facebook according to the number of likes of the posts. A summary of the images is shown in Table 2. In the pan-blue subset, 109 fan pages and 496 posts were included. The number of likes of each post ranged from 904 to 113,168; in pan-green, we included 101 fan pages and 425 posts in the subset. The number of likes of each post ranged from 1,917 to 95,084. We also found DPP (pan-green, the ruling party) uses more images and texts in a post than the opposition (pan-blue).

*Table 2 Top 1,000 images for KMT (pan-blue) and DPP (pan-green) fan page*

| Info | KMT (pan-blue) | DPP (pan-green) |
|---|---|---|
| fan pages | 109 | 101 |
| Posts | 496 | 425 |
| Max numbers of likes | 113,168 | 95,084 |
| Min numbers of likes | 904 | 1,917 |
| Avg. likes/Post | 9,054 | 3,375 |
| Avg. image/Post | 2.01 | 2.35 |
| Avg. text/Image | 19.24 | 26.51 |
| Avg. object/Image | 9.83 | 9.29 |

## Text in Image

We uploaded images to Google vision API, and processed the result as we mentioned in the previous section. After the text segmentation, we removed the stop words in traditional Chinese. Table 3 shows the top 10 most frequent words in pan-blue and pan-green images. We manually merged synonyms; for example, "國民黨" and "中國國民黨" are both represent KMT.

*Table 3 Top 10 most frequent words in pan-blue and DPP (pan-green) images.*

*Table 3 a Top 10 most frequent words in pan-blue images.*

| Top | Words | Frequency |
|---|---|---|
| 1 | Referendum | 326(3.04%) |
| 2 | KMT | 181(1.69%) |
| 3 | Taiwan | 127(1.18%) |
| 4 | Vote | 121(1.13%) |
| 5 | 1218[1] | 88(0.82%) |
| 6 | Disagree | 79(0.73%) |
| 7 | Racto. pork[2] | 66(0.61%) |
| 8 | Government | 64(0.59%) |
| 9 | Health | 60(0.56%) |
| 10 | Algal reef | 52(0.48%) |

*Table 3 b Top 10 most frequent words in DPP (pan-green) images.*

| Top | Words | Frequency |
|---|---|---|
| 1 | Referendum | 959(5.58%) |
| 2 | Disagree | 477(2.77%) |
| 3 | 4 disagree | 391(2.27%) |
| 4 | 1218[1] | 372(2.16%) |
| 5 | TW powerful[3] | 330(1.92%) |
| 6 | Taiwan | 265(1.54%) |
| 7 | Fourth NPP[4] | 231(1.34%) |
| 8 | NGT[5] | 202(1.17%) |
| 9 | Vote | 172(1.0%) |
| 10 | Pork from the US | 166(0.76%) |

[1] Date of the referendum

[2] Pork containing ractopamine

[3] Slogan: Taiwan more powerful

[4] The Fourth Nuclear Power Plant

[5] Natural Gas Terminal

*Table 4 Presentation types of the top 1,000 images in KMT (pan-blue) and DPP (pan-green).*

| Type | KMT (pan-blue) | DPP (pan-green) |
|---|---|---|
| Infographic | 26 | 50 |
| Propaganda | 398 | 561 |
| Others | 576 | 389 |

### Detection on Image

*We retrieved the labels in images through Google vision API. There were 640 labels in 1,000 images from pan-blue fan pages and 541 labels in 1,000 images from pan-green fan pages. We set the threshold at 20 to reserve frequent labels. As shown in Table 5,*

Table 6 and Table 7, we grouped labels into three categories, clothing, physical object, and background.

*Table 5 Clothing labels frequency.*

| Label | KMT (pan-blue) | DPP (pan-green) |
|---|---|---|
| Suit | 16.1% | 29.1% |
| Eyewear | 23.5% | 11.5% |
| T-shirt | 17.5% | 13.2% |
| Trousers | 17.1% | 11.5% |
| Hat | 12.7% | 6.7% |
| Shorts | 3.2% | 2.5% |

*Table 6 Physical object labels frequency.*

| Label | KMT (pan-blue) | DPP (pan-green) |
|---|---|---|
| Poster | 15.2% | 18.7% |
| Tree | 13.6% | 8.2% |
| Vehicle | 20.8% | 7.7% |
| Signage | 6.0% | 6.1% |
| Publication | – | 8.4% |
| Curtain | 3.8% | – |

*Table 7 Background labels frequency*

| | Label | KMT (pan-blue) | DPP (pan-green) |
|---|---|---|---|
| 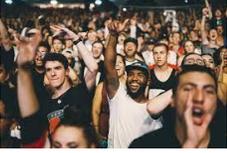 | Crowd | 12.2% | 10.5% |
| 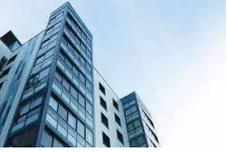 | Building | 8.7% | 3.0% |
| 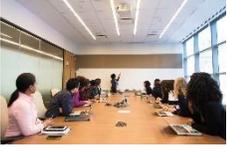 | Community | 5.7% | 4.6% |
| 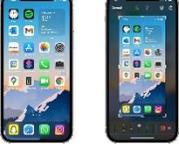 | Screenshot | 4.7% | 6.2% |

## Color Quantization of Image

In this part, we analyzed the color distribution of the top 1,000 images from each group, as shown in Figure 4. We demonstrated the color usage using hue in the HSV system, to better present the distribution of the continuous hue value, we shifted the hue values ranging from 0-50 to 181-231, as shown in Figure 5. The figure shows the polarization of color usage between the two political groups.

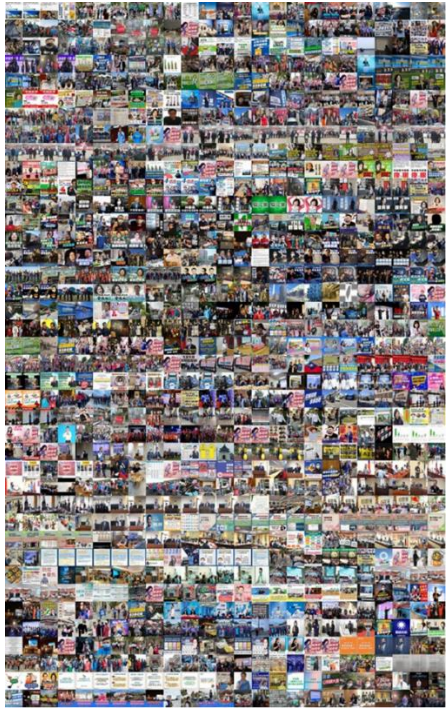 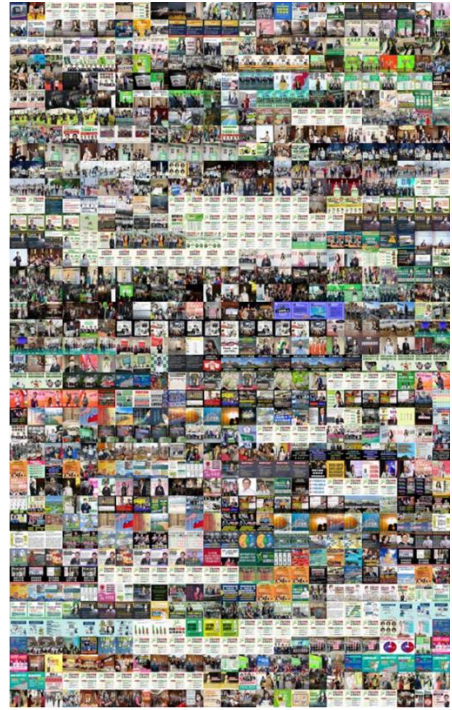

(a)KMT(pan-blue) fan page  (b)DPP(pan-green) fan page

*Figure 4 Top 1000 images per political tendencies fan page.*

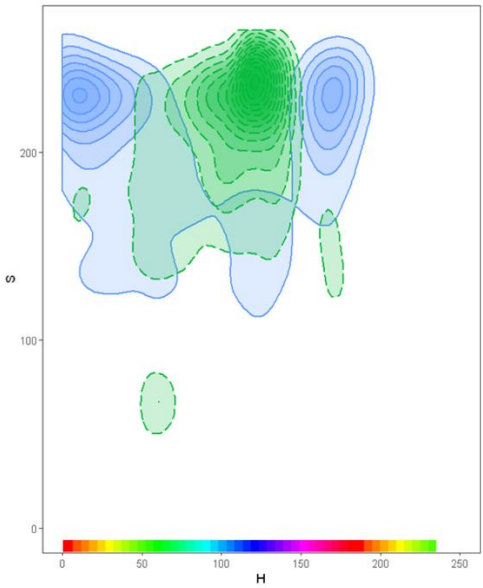 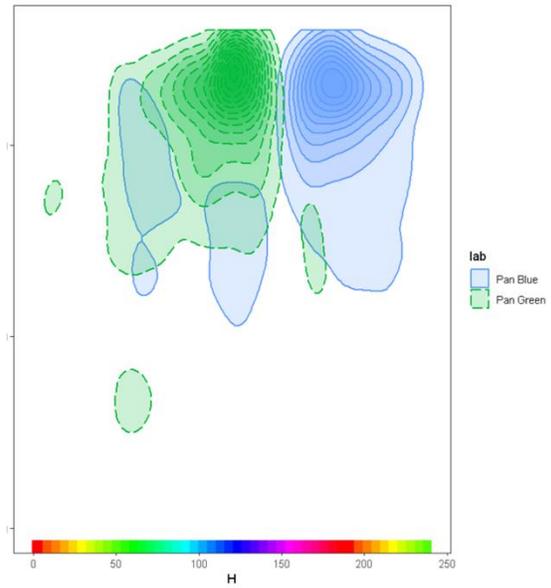

(a)Original hue value distribution  (b)Shifted distribution

*Figure 5 2D density plot for 1000 images.*

## Strategical Analysis

This section described findings from text recognition, object extraction, and color quantization on the images published by different political groups. Also, we investigated the strategical differences between the two groups during propaganda. Two major questions we studied are:

1. Is there any difference between infographic images and propaganda images?
2. Is there any difference between images used to consolidate supporters and attack the opposition?

We attempted to realize the two polarized political groups' strategies for creating propaganda materials by answering the two questions.

We classified the collected images into three categories to address the above questions, including infographics, propaganda, and others. As shown in Table 5, there were only 26 infographic images and 398 propaganda images in the pan-blue group. On the other hand, the pan-green group has published 50 infographics and 561 propaganda images.

According to Table 4, we found pan-green group published 92% more (50 vs. 26) infographic images; the pan-green also used slogan-based propaganda materials more than the pan-blue. From the results, compared with the opposition, pan-blue groups, the ruling party (pan-green) used statistical charts and slogan images to justify their stances about rejecting the referendum proposal.

## Text analysis

We started by listing the top 10 most frequent words extracted from pan-blue and pan-green images, as shown in Table 3. The table shows that the ruling party has a more consistent propaganda slogan. For example, they use "不同意 (disagree)" and "四個不同意 (4 disagree)" to urge citizens to vote against the referendum. From a distribution perspective, pan-green is more purposeful and organized than pan-blue by using more consistent wordings (the top 5 words account for 14.7% of total word usage; pan-blue: 7.9%). The results also show that the two parties are on the opposite side of the issue, such as "四個都同意 (4 agree)" versus "四個不同意 (4 disagree)" and "反萊豬 (oppose pork containing ractopamine)" versus "美豬 (pork from the US)".

We also compared the text extracted from infographics and propaganda images. The significant difference between pan-green and pan-blue groups are the top frequent words in propaganda images. Pan-green fan pages used significant keywords of the referendum, such as "核四 (the Fourth Nuclear Power Plant)"(1.45%), "三接 (Natural Gas Terminal)"(1.36%), and "美豬 (pork from the US)"(1.20%), in the propaganda images; however, none of these words were in pan-blue's top frequent words. From the results, the materials of the government focused on slogans and questions of the referendum; in contrast, the opposition emphasized the political parties and politicians. The reason could be that the referendum can be considered a vote of no confidence in the government. Thus, the ruling party should justify their policies instead of urge votes for certain politicians.

## Object Extraction

*Table 5,*

Table 6, and Table 7 shows the number of clothing, physical objects and backgrounds extracted from images respectively. From these tables, in terms of clothing, pan-blue groups often chose glasses as wear to create a professional image; pan-green groups preferred to wear formal clothing such as suits. The differences between casual wearings such as t-shirts and trousers also demonstrated that pan-blue groups like to wear t-shirts and trousers more than pan-greens. As a party with a history of over 100 years, KMT (pan-blue) politicians seek votes from younger people. In the physical objects, both the numbers of posters appearing in the two parties' images are in the first place, indicating that the political groups used posters in images as the primary means to convey concepts. From Table 7, we also found that pan-blue images include more outdoor scenes are than pan-green's.

## Color Quantization of Image

From the results shown in Figure 5, we observe that both pan-blue and pan-green fan page have their concentrated colors in terms of hue values. As shown in Figure 6, the empirical cumulative distribution function(ECDF) can clearly understand the detail of the concentrated hues. For both pan-blue and pan-green fan pages, there was a phenomenon that they reduced to using the color that high related to their political party. This result is very different from public color perception in different political tendencies; we infer the result was due to the national referendum being for policy debate, but not personal or party.

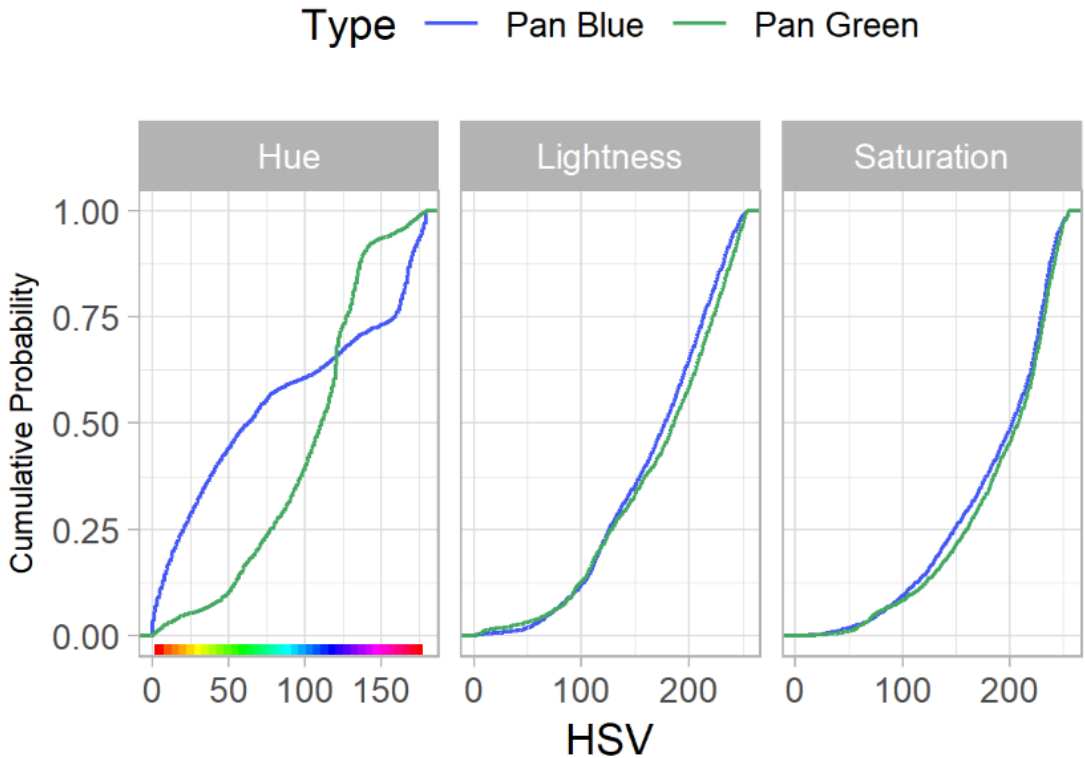

*Figure 6 The ECDF of hue value from 1,000 images of pan-blue and pan-green groups*

### Difference between attacking and consolidation images

To understand how different political groups' fan page designed their images for various purposes, we manually tagged images into the attack and consolidated them as shown in Table 8. We showed some examples of attack and consolidated images for two political groups in Figure 7. After manually tagging images, we investigated the color usage of the two types of pictures, as shown in Figure 8. From the bottom panel of the figure, when consolidating voters, both sides used their representative color more than the opposite; for example, pan-blue fan pages use a more blue tone than pan-green pages. Interestingly, when attacking the opposite, both sides turned to using the opposite's representative color to design their attack images. From the top panel of Figure 8, we observed pan-blue pages used more green tone in attacking images (hue value: 40-60), while pan-green pages used a blue style to attack KMT (hue: 100-120). The results revealed that though the two sides' stances and audiences are polarized, the strategy is similar while attacking/consolidating propaganda.

*Table 8 Different political tendencies images are designed with different purposes.*

| Type | KMT (pan-blue) | DPP (pan-green) |
|---|---|---|
| Attack | 129 | 106 |
| Consolidate | 306 | 457 |

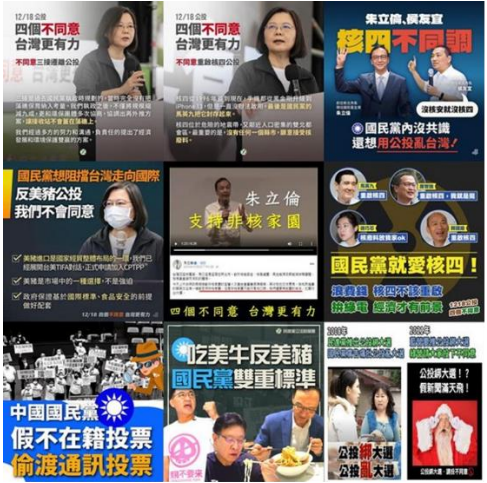
(a) DPP (pan-green) Attack

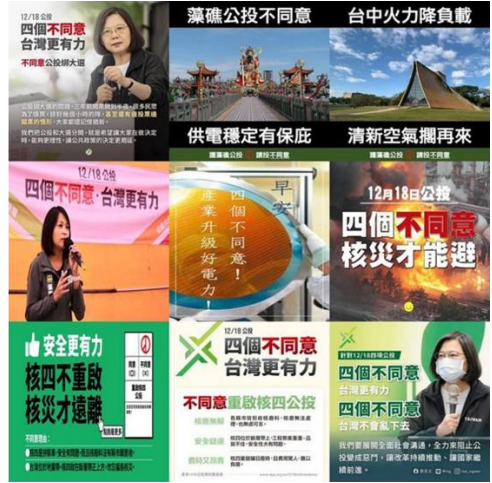
(b) DPP (pan-green) Consolidate

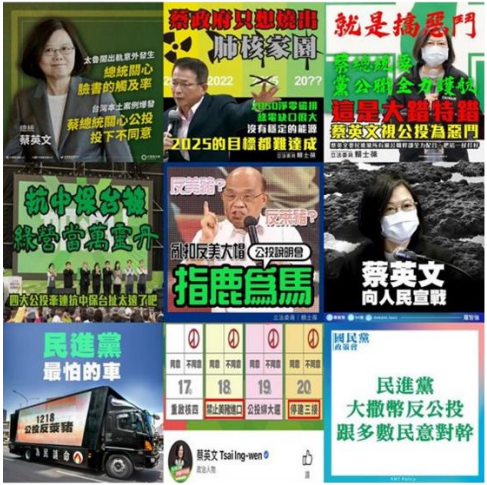
(c) KMT (pan-blue) Attack

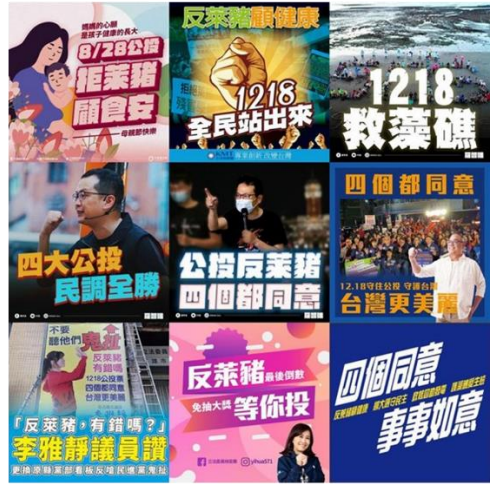
(d) KMT (pan-blue) Consolidate

*Figure 7 The example images about Attack and Consolidate.*

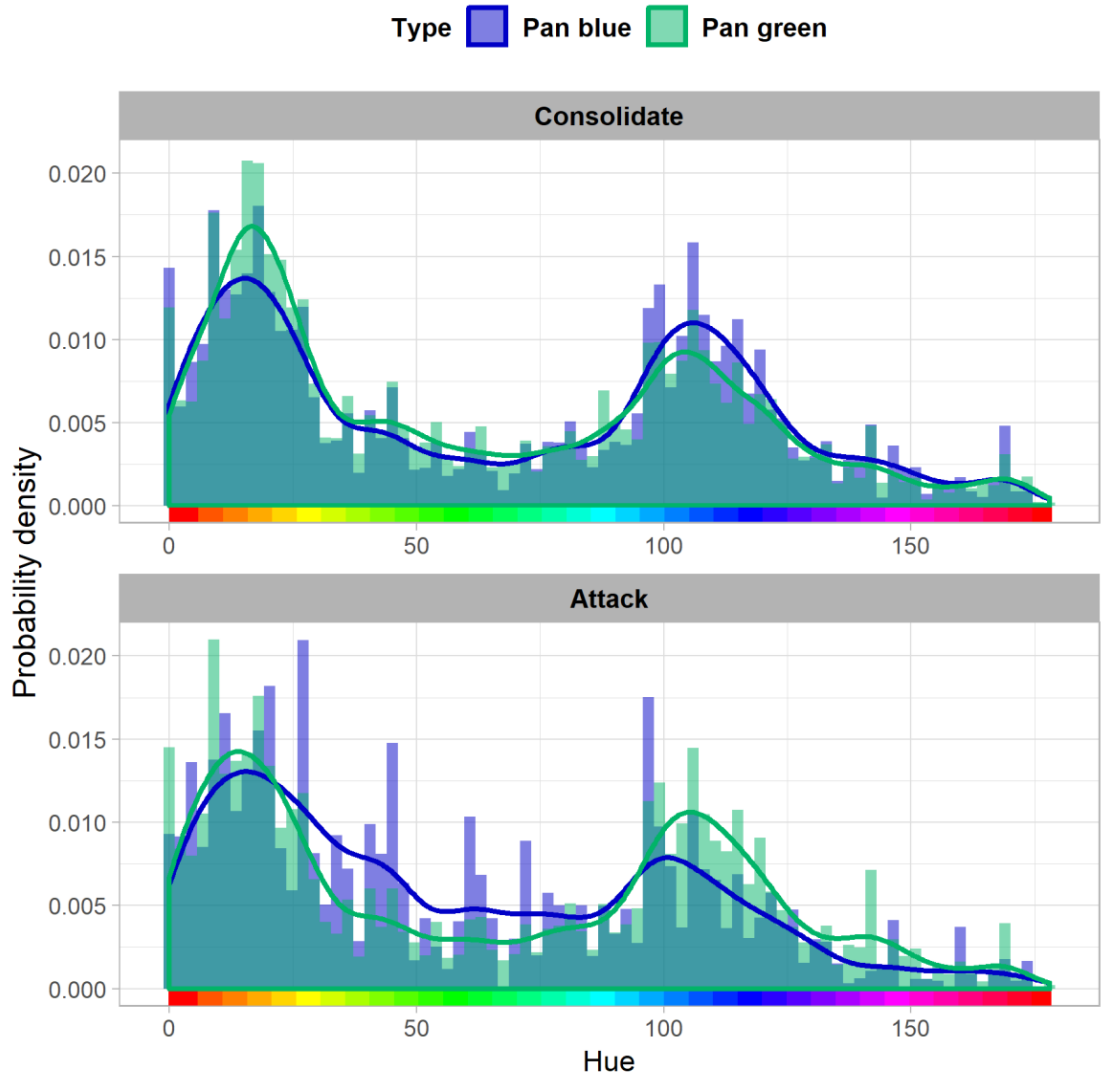

*Figure 8 Compute the empirical cumulative distribution of Attack and Consolidate images for both political tendencies.*

## Conclusion and Future Works

In this paper, we demonstrated the very first study on image-based political communication during a national referendum. Through collecting 1,000 images during the referendum campaign from two polarized political groups' fan pages on Facebook, we analyzed the pictures in terms of texts, objects, and coloring. Compared to the previous studies, we presented significant findings: 1) we found that in a referendum, the government (pan-green in this study) like to use informative images such as infographic and slogan-based images to defend their policy. 2) By extracting objects from images, we found that the century-old party

(KMT, pan-blue) preferred to wear casual clothing, while the ruling party (DPP, pan-green) liked formal wearing. 3) In coloring, our study shows that political groups do not generally emphasize their representative colors, unlike candidate-based elections. A possible reason could be the referendum is for a policy debate but not for individuals or parties. 4) From images with particular purposes, i.e., consolidating or attacking, we found the coloring strategy of both parties in the attack and consolidate images is similar, using their representative color more in consolidating voters and using the opposite's color for attacking.

In future studies, we believe the propaganda strategy will vary along with the development of the events and campaigns. We listed three potential research directions as follows:

- Integrating color psychology and political science to further investigate about which kinds of images could maximize the utility of online propaganda.

- Correlating developing events or poll results to study how political groups react to breaking events and poll results, such as using radical words or representative color to consolidate supporters.

- Studying the similarity and difference between propaganda strategy of the referendum and election campaigns.

We hope this research could shed light on our understanding of online political communication. Moreover, through the quantitative perspectives in this work, we expect this work can facilitate netizens and the public to have novel viewpoints to read, interpret, and react to political propaganda.

## Availability of data and materials

The image dataset is currently available upon request to the corresponding author.

## Funding

This work was supported by Ministry of Science and Technology, Taiwan, under the Grant MOST 111-2622-E-194-005, MOST 110-2927-I-194-001; This work was also supported by Information Operations Research Group, Taiwan.

## Competing interests

The authors declare that they have no competing interests.

## Author's contributions

Conceptualisation: MHW; Data curation: WYC, KHK; Formal analysis and Methodology: MHW, WYC; Visualization: MHW, WYC; First Draft: MHW, WYC; Revisions: MHW, WYC, KHK, KYT; Supervision: MHW, KYT. All authors read and approved the final manuscript.

## Acknowledgements

We would like to thank Information Operations Research Group (https://iorg.tw) for their kind supports in data collection and valuable advice to our work.